\documentclass[journal,twocolumn]{IEEEtran}
\usepackage{amsmath,bm}
\usepackage{amssymb}
\usepackage{latexsym}
\usepackage{multicol}
\usepackage{amsfonts}
\usepackage{graphicx}
\usepackage{subfigure}
\usepackage{float}
\usepackage{color}
\usepackage{cite}
\usepackage{multirow}
\usepackage{diagbox}
\usepackage{algorithm}
\usepackage{algorithmic}
\begin{document}

\title{Image Denoising by Random Interpolation Average with Low-Rank Matrix Approximation}
\author{Qi Liu,
				~Wing-Shan Tam, ~\IEEEmembership{Senior Member, IEEE},
        ~Chi-Wah Kok, ~\IEEEmembership{Senior Member, IEEE},
        \\and~Hing Cheung So,~\IEEEmembership{Fellow, IEEE}
%
\thanks{Q. Liu is with the Department of Electrical and Computer Engineering, National University of Singapore, Singapore (e-mail: elelqi@nus.edu.sg).}
\thanks{H.C. So is with the Department of Electrical Engineering, City University of Hong Kong, Hong Kong, China (e-mail: hcso@ee.cityu.edu.hk).}
\thanks{C.-W. Kok and W.-S. Tam are with the Canaan Semiconductor Limited, Fotan, N.T., Hong Kong.(e-mail: eekok@ieee.org; wstam@ieee.org).}}
\maketitle

\begin{abstract}
With the wide deployment of digital image capturing equipment, the need of denoising to produce a crystal clear image from noisy capture environment has become indispensable. This work presents a novel image denoising method that can tackle both impulsive noise, such as salt and pepper noise (SAPN), and additive white Gaussian noise (AWGN), such as hot carrier noise from CMOS sensor, at the same time. We propose to use low-rank matrix approximation to form the basic denoising framework, as it has the advantage of preserving the spatial integrity of the image. To mitigate the SAPN, the original noise corrupted image is randomly sampled to produce sampled image sets.  Low-rank matrix factorization method (LRMF) via alternating minimization denoising method is applied to all sampled images, and the resultant images are fused together via a wavelet fusion with hard threshold denoising. Since the sampled image sets have independent but identical noise property, the wavelet fusion serves as the effective mean to remove the AWGN, while the LRMF method suppress the SAPN. Simulation results are presented which vividly show the denoised images obtained by the proposed method can achieve crystal clear image with strong structural integrity and showing good performance in both subjective and objective metrics.
\end{abstract}


\IEEEpeerreviewmaketitle

\section{INTRODUCTION}

\IEEEPARstart{I}{mage} denoising is a fundamental and important problem for image processing and computer vision \cite{ImageReview1, ImageReview2, L0-TV, ImageReview}. It is because the natural image is inevitably contaminated by noise during phases of acquisition and transmission, which is the major source of noise degrading the image quality in the subsequent image processing application, such as object segmentation, edge detection, feature extraction, etc \cite{4079662, 5430907, SELF}.

Many techniques have been proposed to perform image denoising, e.g., spatial domain-based methods \cite{NLM}, statistical modeling based method \cite{MKM199}, order statistics method \cite{donoho95}, and transform domain-based method \cite{BM3D,PLOW, SALSA, CSR, RIA1}, etc. The transform domain methods are popular in literatures because of its effectiveness in reducing the complexity of the image representation, where they first transform the spatial domain image into transform domain image. A natural image can be considered as a locally stationary Gaussian process which contains a self-similarity features plus noise. Therefore, it can be represented by a set of sparsely decomposed coefficients or known as a dictionary of atoms, such as the set of cosine functions in Discrete Cosine Transform (DCT), or the wavelet bases in wavelet transform, and fractal codes in Partitioned Fractal System (PFS) \cite{5420029, 5466111, CSR}. With the sparsity in the dictionary, a noise-free image can be estimated by discarding the redundancy in the dictionary.
In \cite{4011956, K-SVD2, K-SVD1}, the authors proposed to use K-SVD \cite{K-SVD} to learn the dictionary from noise-free image set. Due to its good performance, many methods based on sparse representation and dictionary learning have been extended to nonlocal models, which include the locally learned dictionaries (K-LLD) \cite{K-LLD}, clustering-based sparse representation (CSR) \cite{CSR} and Robust kronecker component analysis (RKCA) \cite{RKCA}. However, their application is limited by the high computational cost  in dictionary learning. In contrast to learning a global dictionary or dictionaries for each patch-cluster, the proposed HOSVD in \cite{HOSVD} learns bases that vary from pixel to pixel, obviating the need for any iterative optimization. In \cite{SALSA}, although the split augmented Lagrangian shrinkage algorithm  (SALSA) applied in small noise image has achieved a good performance and faster computation, its result is unsatisfactory especially in strong additive white Gaussian noise (AWGN). In addition, the performances of these sparse denoising methods are often limited by the sparsifying capacity of the transform, such that the more sparse the representation, the better the denoising result. Nevertheless, since these methods need to decompose the noisy image into several blocks (patches) first, and then perform sparse representation for each small block under certain fixed atom library, the resultant sparsity optimization problem with dictionary learning is still computationally demanding \cite{BM3D}. What is more, some of them need to know the noise variance to normalize the F-norm data fidelity term so as to denoise the image, which is not practical since it may be hard to detect the noise variance especially in the case of the mixed noise.

Another class of the image denoising scheme is  the spatial domain technique, which directly tackles the intensity of each pixel of the image, such as bilateral filter \cite{BF}, non-local means (NLM) filter \cite{NLM} and guided image filter. Due to the limitation of bilateral filter in preserving gradient direction of edges, NLM filter has been considered to have improved the overall denoising performance. Recently, the emerging technique of low-rank matrix approximation (LRMA) has given renewed interest to image denoising, and has a wide range of applications in computer vision. In weighted nuclear norm minimization (WNNM) algorithm \cite{WNNM}, the low-rank regularization is enforced to reconstruct the latent structure of the noisy patch matrix. However, it only considers the non-local self-similarity property of the noise corrupted image, which makes it difficult to denoise the image from the noisy observation alone.  

In the view of aforementioned difficulties, we propose a new image denoising method to tackle both salt-and-pepper noise (SAPN) and AWGN, which takes full advantages of the two-dimensional version of the random interpolation averaging (2-D RIA) method with good performance and robustness against impulsive noise. The LRMF method via alternating minimization (denoted as LRMF-AM) is applied to the 2-D RIA generated image set. The processed images are fused together via wavelet fusion with hard threshold denoising to alleviate the AWGN in the LRMF-AM image set.

The contributions of this paper can be summarized as follows. First, the RIA scheme \cite{RIA, RIA1} has been successfully extended to 2-D RIA image denoising with the help of the LRMF-AM to mitigate the mixed noise (SAPN and AWGN). The residual AWGN is also further alleviated by the proposed wavelet fusion with hard thresholding on the denoised 2-D RIA image set. Simulation results show that the denoising results can achieve better performance under both quantitative and visual quality measures when compared with existing methods in the literature.

The rest of this paper is organized as follows. In Section II, the image modelling in spatial domain is defined. In Section III, image denoising by means of low-rank matrix and LRMF denoising method is formulated and is applied to the image set obtained from applying 2-D RIA with the noise corrupted image. A wavelet fusion with hard threshold method is proposed in Section IV to alleviate the residual AWGN in images obtained from LRMF-AM denoising in the 2-D RIA image set. The fused image is the final denoised image, which is demonstrated to have superior clarity when compared to existing techniques as shown by the simulation results in Section V. Finally, conclusions are drawn in Section VI.

\section{Image Modelling}

There are two major noise sources that affect the quality of digital images, which are the AWGN due to the random noise nature of the image capturing device; and the SAPN due to hot pixels caused by current leakage in CMOS image sensors. Denoising an image $\mathbf M \in \mathbb{R}^{M\times N}$ corrupted with mixed AWGN and SAPN to obtain a noise-free image $\mathbf X$ is a challenging task. Without loss of generality, the AWGN and SAPN are assumed independent, such that the realistic model is written as \cite{Cai2008TwophaseAF, Cai2009}:
\begin{eqnarray}\label{Y}
	\mathbf Y &=& \mathbf X + \mathbf N +\mathbf S,\\
	\mathbf M &=& Q(\mathbf Y),
\end{eqnarray}
where $\mathbf{N}[i,j] \in\mathbb Z$ denotes the zero-mean AWGN, and $\mathbf{S}[i,j] \in[-\mbox{max},\mbox{max}]$ denotes a sparse matrix composed in the set of integer of impulsive noise which takes value as either the maximum available pixel brightness $\mbox{max}$ (the salt) or the negative of it, such that the final pixel value equals 0 (the pepper). The entities in in $\mathbf N$ have the values spanning through the dynamic range of the pixel intensity in the image, while $\mathbf S$ has its entities values at either the maximum and minimum values of dynamic range to indicate the salt and pepper noises, respectively. The quantizer $Q$ is defined as:
\begin{eqnarray}
	Q(x) = \left\{{\begin{array}{ll} x & 0\leq x\leq \mbox{max}, \label{quantizer}\\
	\mbox{max} & x>\mbox{max}, \\
	0 & x<0. \end{array}}\right. 
\end{eqnarray}
For gray-level image with 8 bits per pixel, the dynamic range of the pixel is  $[0,255]$, with $\mbox{max}=255$. In actual, the quantizer also represents an impulsive noise which take their values in the dynamic range $[0, 255]$, known as the random-valued noise \cite{Cai2008TwophaseAF, Cai2009}. Herein, based on the definition of the quantizer, the random-valued noise is regarded as SAPN. The quantizer can be taken out from $\mathbf M$ by considering the AWGN corrupted image $\mathbf P$ at pixel location $[i,j]$ is given by
\begin{eqnarray}
	\mathbf P = \mathbf X +\mathbf N,
\end{eqnarray}
and the SAPN on the AWGN corrupted image 
\begin{eqnarray}
	{\mathbf S}[i,j] = \left\{\begin{array}{ll}\mbox{max}-{\mathbf X}[i,j] & \mbox{with probability $p_1$,}\\
	-{\mathbf X}[i,j] & \mbox{with probability $p_2$,}\\
	{\mathbf X}[i,j] & \mbox{with probability $1-p_1-p_2$.}\end{array}\right.
\end{eqnarray}
As a result, the SAPN corrupted image $\bf M$ with AWGN corrupted image $\mathbf P$ is rewritten as:
\begin{eqnarray}\label{modelM}
	\mathbf M &=& Q((\mathbf X + \mathbf N) + \mathbf S)\notag\\
	&=& Q(\mathbf P + \mathbf S).
\end{eqnarray}
When the SAPN power is low, $\mathbf S$ will be a sparse matrix of impulsive noise. From model \eqref{modelM}, we see that even when we recover the image from the SAPN corrupted image $\mathbf M$, the estimated image $\mathbf P$ is still contaminated by the AWGN, which makes the procedure of denoising AWGN corrupted image in the following Part IV reasonable. This model without the quantizer has been considered in \cite{Zeng2018} on an inpainting problem, where the underlying image is modeled to have a minimum total variation. For the noisy image $\mathbf M$ with $\mathbf M[i,j] \in [0,255]$, it is relatively hard to detect the SAPN pixels for those mixed noise removal methods \cite{Cai2008TwophaseAF, Cai2009, 1621237, XIAO20111708}, where they have to perform impulsive noise pixel detection at first. The reason is that the quantizer has mixed those salt or pepper noise pixels with noise-free image pixels being 0 or 255. One natural question is that can we develop a mixed noise removal method which does not perform SAPN pixel detection?

The LRMA aims to recover the underlying low-rank matrix from its degraded observation, which can be generally achieved by regularization based models \cite{7959547, WSNM, 7062902} and factorization based models \cite{TLiu2016, VBMFL1, AMB2005} . One of the most representative low-rank regularizers in regularization based models is the nuclear norm\footnote{The nuclear norm is to compute the sum of singular values} to relax the highly discrete and nonconvex rank, which is analogous to the strategy of employing the $\ell_1$-norm instead of the $\ell_0$-norm for sparse signal recovery \cite{4472240}.  It is concluded that the drawbacks of using nuclear norm, however, are twofold: since the resulting nuclear norm minimization can be converted into a semidefinite program (SDP) in \cite{SDP}, it can be solved by the interior-point methods, including singular value thresholding \cite{SVT}, fixed point continuation \cite{FPC}, and proximal gradient descent \cite{PGD}, etc., these algorithms with nuclear norm relaxation have to perform whole matrix SVD at each iteration. When the matrix size is large, they have high computational cost; Second, because of all the singular values being dealt with equally, the larger singular values are penalized more heavily, meaning that the nuclear norm is not a satisfactory surrogate of the rank function in real applications. Thus, the LRMA with regularization based models has to choose the optimal regularization parameter in iterations. Therefore, to avoid the above drawbacks, we focus on the latter category in this work.

As a result, under the assumption of low-rank, SAPN problem in our work is formulated as the following optimization problem:
\begin{eqnarray}\label{norm:min:prob}
	\min_{\mathbf A, \mathbf B}\| (\mathbf A \mathbf B) - \mathbf M\|^2_F,
\end{eqnarray}
where $\mathbf A \in \mathbb{R}^{M \times r}$ and $\mathbf B \in \mathbb{R}^{r \times N}$ with $r \ll{\rm min}\{M, N\}$, and $\|\cdot\|_F$  denotes the Frobenius norm. 

In general, the existing denoising works address the SAPN problem with $\ell_p$-norm\footnote{The $\ell_p$-norm of a matrix denotes $||\mathbf P||_p = (\sum_{i, j} |P_{i, j}|^p)^{1/p}$} minimization of the error term, where they design the denoising algorithms using $\ell_p$-norm to restore the noise-free image from the noisy measurements. However, some of them cannot guarantee its convergence due to the nonconvexity of the resulting problem with highly nonconvex $\ell_p$-norm. Moreover, the denoising performance of them is greatly affected by the choice of $p$. It is well known that least squares-based metric is highly sensitive to outliers present in the measurement vectors, leading to poor recovery. In this work, the reasons using Frobenius norm data fitting model as the metric in \eqref{norm:min:prob} are included as follows. First, the goal of LRMA exploited in \eqref{norm:min:prob} is not for the SAPN reduction, instead of the randomly sampled image reconstruction for 2-D RIA. Any signal reconstruction method from randomly sampled image can be applied. Herein, the resultant Frobenius norm minimization is utilized due to its convexity. Thus, since the Frobenius norm is convex in the reconstruction procedure, it is easier to find the closed-form solution based on the resultant least squares problem. Second, the idea of the proposed scheme is to utilize the random sampling and multiple images averaging to denoise SAPN, where the better denoisng performance from the following simulation results also implies that the reasonability of using Frobenius norm and the simple implementation of our scheme with Frobenius norm. The reason of better denoising performance is that the implemented method does not rely on the use of straight eigen-image, thus it provides relaxation to preserve complex image features when compared to that using few eigen-images alone to approximate the original image. Finally, the AWGN is being removed by averaging multiple SAPN removed images and we implement the averaging is a way that perform wavelet denoising to ensure the best reduction of AWGN. Since the key on the removal of SAPN in the proposed scheme depends on the random sampling and multiple image averaging, the use of Frobenius norm or $\ell_p$-norm formulations would affect the denoising performance but it would not be critical.

In the following sections, we shall describe how to make use of the above cost function to denoise a SAPN corrupted image with 2-D RIA, which will yield a denoised image with AWGN residual noise. A separate denoising scheme will then be applied to denoise the AWGN corrupted image to yield superior denoising results. 

\section{Image Modeling as Low-Rank Matrix}

Our work directly addresses \eqref{norm:min:prob} to reconstruct the random sampled images for 2-D RIA. The LRMF method \cite{LRMCam, SFM} has been exploited for the norm minimization problem in \eqref{norm:min:prob}, where the unknown matrix is modeled as a product of two matrices with much smaller dimensions so that the low-rank property is automatically fulfilled. Herein, the unknown matrix $\mathbf P$ is factorized into two smaller matrices $\mathbf A \in \mathbb{R}^{M \times r}$ and $\mathbf B \in \mathbb{R}^{r \times N}$. Therefore, for each random sampled image, \eqref{norm:min:prob} is converted to the following matrix Frobenius norm minimization:
\begin{align}
\label{bi-convex-problem}
\min_{\mathbf A, \mathbf B} f(\mathbf A, \mathbf B) \triangleq ||(\mathbf A \mathbf B) - \mathbf M_\ell||_F^2, \ell = 1, \cdots, k
\end{align}
where $f(\mathbf A, \mathbf B)$ is defined as the error function, and $k$ is the number of samples in two image sets. Unfortunately, the matrix Frobenius norm minimization problem in \eqref{bi-convex-problem} with respect to (w.r.t.) $\mathbf A$ and $\mathbf B$ is a nonconvex optimization problem and is therefore intractable. However, it is worth noting that if we fix one of the two matrices,  $\mathbf A$ or $\mathbf B$, the error function $f(\mathbf A, \mathbf B)$ is convex w.r.t. the free matrix and the global minimum is readily available\footnote{ It is noticed that the alternating optimization procedure does not necessarily converge to the global minimum point. The point that it converges to depends on the initial value $\{\mathbf A^{(0)}, \mathbf B^{(0)}\}$.}. As a result, an alternating minimization  \cite{LRMCam,bi-convexProblem} is applied to solve the resultant bi-convex optimization problem in \eqref{bi-convex-problem}. Consider $\mathbf A^t$ and $\mathbf B^t$ at the $t$th iteration,  the matrices $\mathbf A^{t+1}$ and $\mathbf B^{t+1}$ are obtained by the following two optimization problems sequentially: 
\begin{align}
\label{R-sub-problem}
\min_{\mathbf B} f(\mathbf B^{t+1}) = ||(\mathbf A^t \mathbf B) - \mathbf M_\ell||_F^2,
\end{align}
\begin{align}
\label{Q-sub-problem}
\min_{\mathbf A} f(\mathbf A^{t+1}) = ||(\mathbf A \mathbf B^{t+1}) - \mathbf M_\ell||_F^2.
\end{align}
Both \eqref{R-sub-problem} and \eqref{Q-sub-problem} are convex and it is obvious that they have the same structure and can be solved in a similar manner. Therefore,  without loss of generality, we shall discuss the solver for $\mathbf B$, while the same method can be applied to $\mathbf A$ with straightforward modifications. Denote $\mathbf A = [\mathbf a_1, \cdots, \mathbf a_M]^T, \mathbf B = [\mathbf b_1, \cdots, \mathbf b_N]$, where the $i$th row of $\mathbf A$ and the $j$th column of $\mathbf B$ are represented as $\mathbf a_i^T \in \mathbb{R}^r, i = 1, ..., M$ and $\mathbf b_j \in \mathbb{R}^r, j = 1, ..., N$, respectively. For notational simplicity, the superscript $(\cdot)^t$ that denotes the iteration number  is dropped in the following analysis without affecting the solution. To solve \eqref{R-sub-problem} for $\mathbf B$ with a fixed $\mathbf A$ \cite{LRMCam}, we  rewrite \eqref{R-sub-problem} as:
\begin{align}
\label{solvingR-fixedQ}
\min_{\mathbf b_j} f(\mathbf b_j) = \sum_{j}|\mathbf A \mathbf b_j - \mathbf m_j|^2,
\end{align}
where $\mathbf m_j$ denotes the $j$th column of $\mathbf M_\ell$. Since $f(\mathbf B)$ is decoupled w.r.t. $\mathbf b_j$, the solution for \eqref{solvingR-fixedQ} can be obtained by finding the solutions of the following $N$ independent sub-problems:
\begin{align}
\label{NV-sub-problem}
\min_{\mathbf b_j} f(\mathbf b_j) = ||\mathbf A \mathbf b_j - \mathbf m_j||_2^2,
\end{align}
which is a least squares problem, and its solution is $\mathbf b_j = \mathbf A^{\dagger}\mathbf m_j $. Similarly, we can apply the same method to find the solution $\mathbf A$ for \eqref{Q-sub-problem}, and its solution is $\mathbf a_i = (\mathbf B^T)^\dagger \mathbf m^T_i$, where $\mathbf m_i$ denotes the $i$th row of $\mathbf M_\ell$. After determining $\mathbf A$ and $\mathbf B$, the target matrix is obtained as $\mathbf {\hat P} = \mathbf A \mathbf B$. Note that the global optimum solution $\{\mathbf A, \mathbf B\}$ for \eqref{bi-convex-problem} is not guaranteed. The reason lies in the fact that for an arbitrary orthogonal matrix $\mathbf T$ which fulfills $\mathbf T \mathbf T^T = \mathbf I$, we have $\mathbf {\hat A} = \mathbf A \mathbf T, \mathbf {\hat B}= \mathbf T^{-1}\mathbf B$ that  are also solutions for \eqref{R-sub-problem} and \eqref{Q-sub-problem}. Fortunately, we are not interested in $\mathbf A$ or $\mathbf B$ itself, while the orthogonal transform $\mathbf T$ does not affect $\mathbf {\hat P}$. Therefore, the solution obtained from \eqref{NV-sub-problem} is still optimal for our application. One advantage of the matrix factorization based approach is that the SVD can be avoided. It is clear that the complexity of the convex optimization problem in \eqref{R-sub-problem} is $\mathcal O(MNr^2)$, which is the same as the complexity of \eqref{Q-sub-problem}. Hence, the computational complexity of the proposed method in each iteration is $\mathcal O(MNr^2)$.

The proposed algorithm (LRMF-AM) in Algorithm \ref{LRMF-AM} does not only has low computational complexity, it also has good convergence. We shall first show that $||\mathbf A \mathbf B - \mathbf M_\ell||_F^2$ converges to a local minimum. Then extend our proof to show that $\{\mathbf A, \mathbf B\}$ will also converge to a local optimum. Let the objective value $||\mathbf A \mathbf B - \mathbf M_\ell||_F^2$ after solving the two sub-problems in \eqref{R-sub-problem} and \eqref{Q-sub-problem} be $E^t_{\mathbf B}$ and $E^t_{\mathbf A}$, respectively, at the $t$th iteration. Since we have $E^t_{\mathbf B} = ||\mathbf A^{t-1} \mathbf B^t - \mathbf M_\ell||_F^2$ and $E^t_{\mathbf A} = ||\mathbf A^{t} \mathbf B^t - \mathbf M_\ell||_F^2$, the local optimality of $\mathbf A^t$ yields $E^t_{\mathbf B} \ge E^t_{\mathbf A}$. Similarly, at the $(t+1)$-th iteration, $E^t_{\mathbf A} = ||\mathbf A^{t} \mathbf B^t - \mathbf M_\ell||_F^2$ and $E^{t+1}_{\mathbf B} = ||\mathbf A^{t} \mathbf B^{t+1} - \mathbf M_\ell||_F^2$, the local optimality of $\mathbf B^{t+1}$ yields  $E^t_{\mathbf A} \ge E^{t+1}_{\mathbf B}$. Therefore, we get $E^1_{\mathbf B} \ge E^1_{\mathbf A} \ge E^2_{\mathbf B} \ge \cdots \ge E^t_{\mathbf B} \ge E^t_{\mathbf A} \ge E^{t+1}_{\mathbf B} \ge \cdots$, that is, the objective value $||\mathbf A \mathbf B - \mathbf M_\ell||_F^2$ does not increase at each iteration and is bounded below. The above bounded norm sequence does not automatically imply the convergence of the elements $\{\mathbf A^t\}$, and $\{\mathbf B^t\}$. On the other hand, the proposed method utilizing alternate sequence optimization is a special case of the block coordinate descent algorithm. Both \eqref{R-sub-problem} and \eqref{Q-sub-problem} are Gateaux-differentiable w.r.t. the corresponding variables over its open domain. Using the cyclic rule, each coordinate-wise minimum point of the proposed method is a stationary point according to Theorem 5.1 in \cite{BCD}. Therefore, the alternating optimization in \eqref{R-sub-problem} and \eqref{Q-sub-problem} produces a sequence of $||\mathbf A \mathbf B - \mathbf M_\ell||_F^2$ that converges to a stationary point.

\begin{algorithm}[http]
\caption{LRMF-AM}
\label{LRMF-AM}
\begin{algorithmic}
\REQUIRE {$\mathbf M$, ${\mathcal{I}_j}$, ${\mathcal{J}_i}$, $r$}\\
\STATE{{\bf Initialize:} $\mathbf A^0 \in \mathbb{R}^{M\times r}$, $j = 1, \cdots, N$, $i = 1, \cdots, M$}
\FORALL{$\ell = 1, 2, \cdots, k$}
\IF{$t = 0, 1, \cdots $}
\REPEAT
\STATE{$\mathbf {\hat b}_j =$ arg $\min \limits_{\mathbf b_j}$ $||\mathbf A^{t}\mathbf b_j - \mathbf m_j ||^2_2$}
\STATE{$\mathbf {\hat a}_i =$ arg $\min \limits_{\mathbf a_i^T}$ $||\mathbf a^T_i \mathbf B^{t+1}-\mathbf m_i ||^2_2$ \\ ~~ $=$ arg $\min \limits_{\mathbf a_i}$ $||(\mathbf B^{t+1})^T \mathbf a_i -\mathbf m^T_i ||^2_2$}
\UNTIL{termination condition satisfied.}
\ENDIF
\ENDFOR
\ENSURE{$\mathbf {\hat P}_\ell = \mathbf{AB}$}\\ 
\end{algorithmic}
\end{algorithm}

\begin{figure}[htb]
\centering
\subfigure[]{
\begin{minipage}{0.4\linewidth}\centering
\includegraphics[width=2 in]{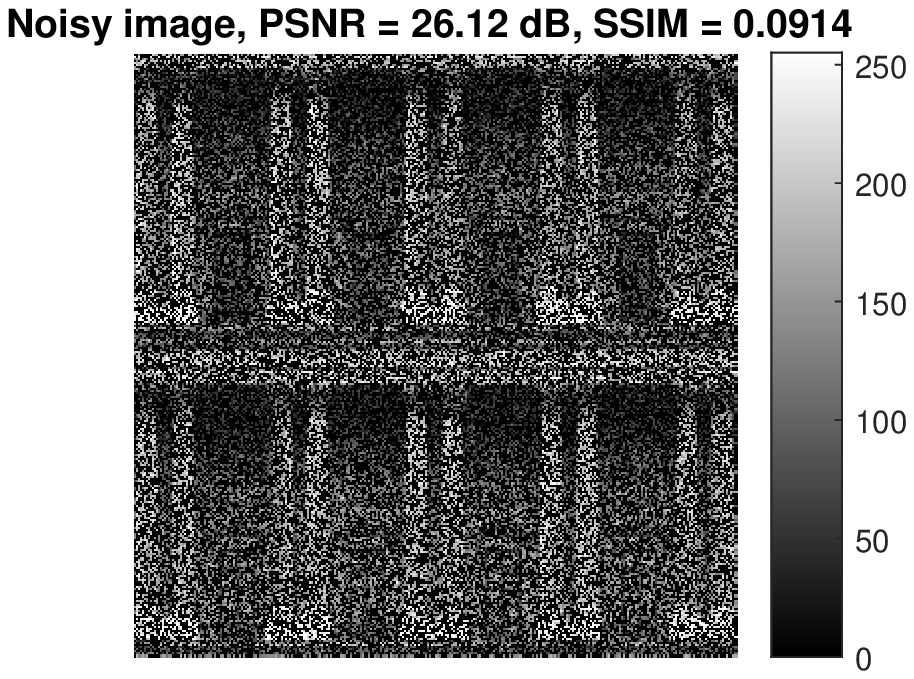} 
\end{minipage}
}
\subfigure[]{
\begin{minipage}{0.5\linewidth}\centering
\includegraphics[width=2 in]{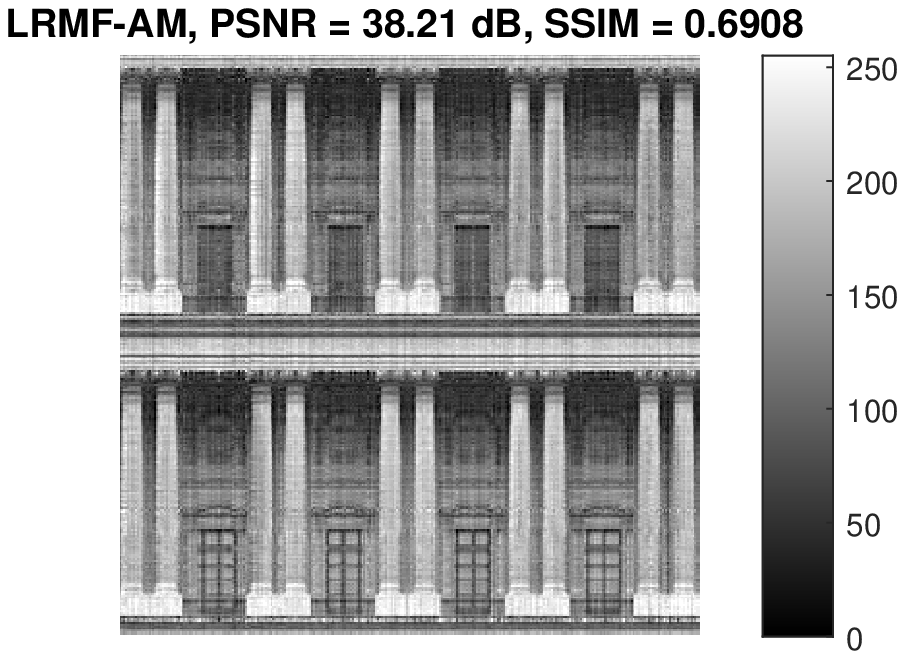} 
\end{minipage}
}

\subfigure[]{
\begin{minipage}{0.4\linewidth}\centering
\includegraphics[width=2 in]{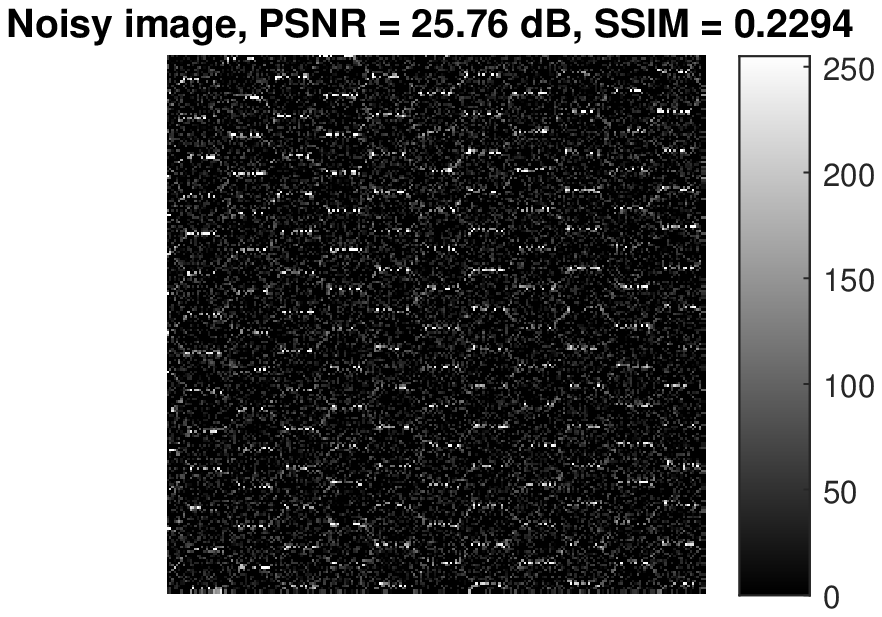}
\end{minipage}
}
\subfigure[]{
\begin{minipage}{0.5\linewidth}\centering
\includegraphics[width=2 in]{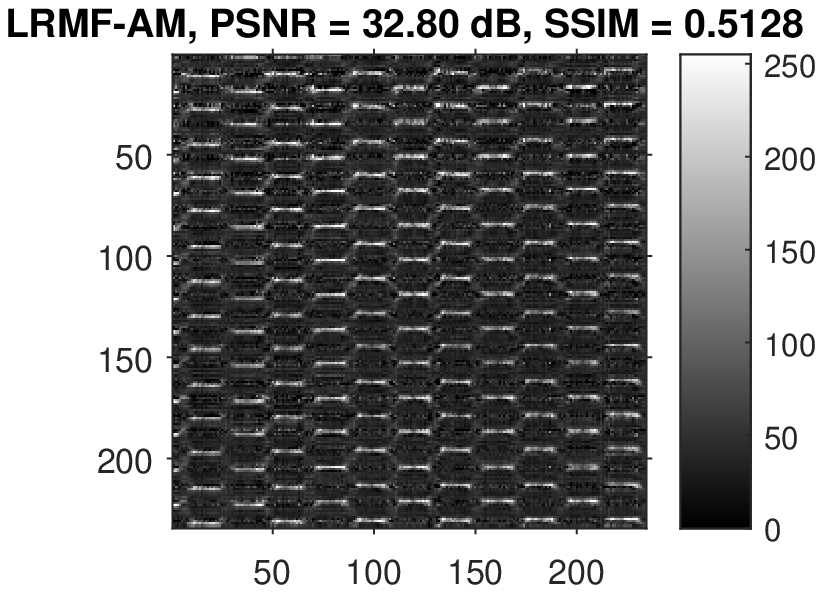}
\end{minipage}
} 
\caption{Denoising performance by LRMF-AM with rank $r = 10$ and $k = 4$. The left column figures are noisy images corrupted by the AWGN and SAPN with $\sigma =20$ and $\rho =0.3$, while the reconstructed images are shown in the right column.}
\label{LRMF1}
\end{figure}

To evaluate the performance of the locally converging LRMF-AM denoising algorithm, consider SAPN ($\rho=p_1+p_2=0.3$) plus AWGN ($\sigma=20$) corrupted images as shown in Figs.~\ref{LRMF1} (a) and (c), where $r = 10$. By applying the LRMF-AM denoising method, the reconstructed images are shown in Fig.~\ref{LRMF1} (b) and (d), respectively, where $\mathbf {\hat P} $ is fused by averaging $\mathbf {\hat P}_\ell$ with $k = 4$. The denoising results are in general good but the reconstructed images still observed to be corrupted by AWGN in the smooth areas, such as along the columns in Fig.~\ref{LRMF1} (b) and the dark background of Fig.~\ref{LRMF1} (d). A separate AWGN denoising procedure that can work hand in hand with the proposed LRMF-AM based SAPN denoising method will be presented in the following section to achieve the final denoising results for SAPN plus AWGN corrupted images. 

\section{Denoising AWGN Corrupted Image}

\begin{figure}[!htbp]
\centering
\subfigure{
\begin{minipage}{0.4\linewidth}\centering
\includegraphics[width=2 in]{L2_50perc_noisy.eps} 
\end{minipage}
}
\subfigure{
\begin{minipage}{0.4\linewidth}\centering
\includegraphics[width=2 in]{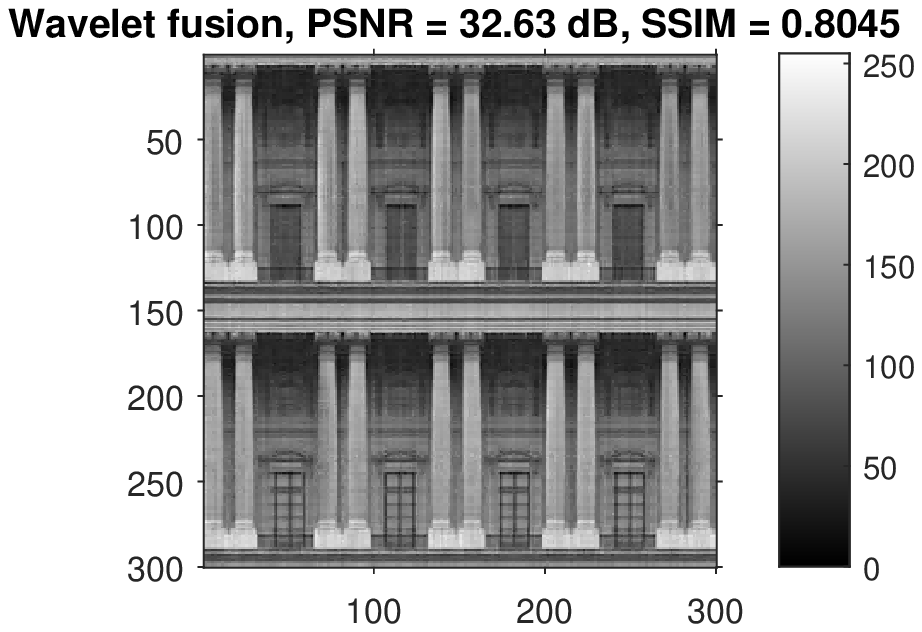} 
\end{minipage}
}

\subfigure{
\begin{minipage}{0.4\linewidth}\centering
\includegraphics[width=2 in]{L2_50perc_text_noisy.eps}
\end{minipage}
}
\subfigure{
\begin{minipage}{0.4\linewidth}\centering
\includegraphics[width=2 in]{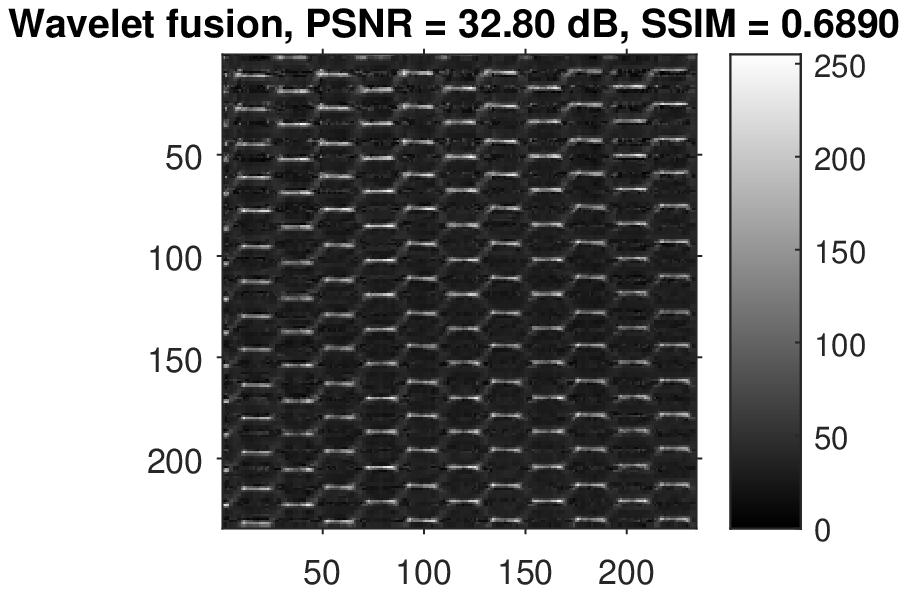}
\end{minipage}
}
\caption{AWGN denoising by image fusion, where $\tau = 15$ and $k = 4$. The left column figures are noisy images, while the right column images are fused by wavelet fusion technique.}
\label{fig:wavelet:fusion}
\end{figure}

The simulation results presented in previous section show that the SAPN denoised image by low-rank image model has effectively alleviated the SAPN corrupted image, but the resulting image is still suffering from residual AWGN. To denoise AWGN corrupted image, we consider the case of the image $\mathbf P$ being corrupted by $k$ different processes of zero-mean AWGN with the same variance and generates a set of noise corrupted images $\mathbf P_\ell$ with $\ell\in 1,2,\ldots,k$. If all $k$ AWGN processes are independent and identically distributed (IID), an effective noise removal routine will be fusing all $k$ AWGN corrupted images together to form a single image. The mean filtering effect of the image fusion process will help to suppress the AWGN because of their independent nature. There exists many different ways to fuse $\mathbf P_\ell$ together. Fig.~\ref{fig:wavelet:fusion} shows an example of fusing two images $\mathbf P_1$ and $\mathbf P_2$ together via wavelet, which we shall demonstrate in subsequent section that this simple method does work perfectly together with LRMF-AM SAPN denoising. The steps of wavelet fusion with hard thresholding are summarized in Algorithm \ref{Wavelet}.

\begin{algorithm}[htb]
\caption{Wavelet Fusion with Hard Thresholding}
\label{Wavelet}
\begin{algorithmic}
\REQUIRE {$\mathbf P_\ell$, $\tau$, $k$}\\
\FOR{$\ell = 1, 2, \cdots, k$}
\STATE {Perform DWT on $\mathbf P_\ell$}\\
\STATE{${\mathbf P_\ell}\stackrel{DWT}{\rightarrow} [\begin{array}{llll}{\mathbf{LL}}_\ell & {\mathbf{LH}}_\ell & {\mathbf{HL}}_\ell & {\mathbf{HH}}_\ell\end{array}]$}
\ENDFOR
\STATE{~~~// Fusing all frequency subbands together}
\STATE{~~~$\mathbf{LL} {\leftarrow}$ \eqref{LL} }
\STATE{~~~$\mathbf{LH}, \mathbf{HL}, \mathbf{HH} {\leftarrow}$  \eqref{LH-HL-HH}}
\STATE{~~~// Perform the IDWT}
\STATE{~~~$\mathbf {\hat X} {\leftarrow}$  \eqref{IDWT}}
\ENSURE{$\mathbf {\hat X}$}\\ 
\end{algorithmic}
\end{algorithm}

The wavelet denoising method presented in \cite{donoho95} suggests  decomposition of the image $\mathbf P_\ell$ into subband images through discrete wavelet transform (DWT).
\begin{eqnarray}
	{\mathbf P_\ell}\stackrel{DWT}{\rightarrow} [\begin{array}{llll}{\mathbf{LL}}_\ell & {\mathbf{LH}}_\ell & {\mathbf{HL}}_\ell & {\mathbf{HH}}_\ell\end{array}]. 
\end{eqnarray}
A hard thresholding method is then applied to the each components of the high frequency subband coefficients with threshold parameter $\tau$ to remove low power noise signal:
\begin{eqnarray}
{\mathbf{LH}}_\ell[i,j] &=& \left\{\begin{array}{ll}0, & |{\mathbf{LH}}_\ell[i,j]|<\tau, \\
             {\mathbf{LH}}_\ell[i,j], & \mbox{otherwise;} \end{array} \right. \nonumber \\
{\mathbf{HL}}_\ell[i,j] &=& \left\{\begin{array}{ll}0, & |{\mathbf{HL}}_\ell[i,j]|<\tau, \\
             {\mathbf{HL}}_\ell[i,j], & \mbox{otherwise;} \end{array} \right. \nonumber \\
{\mathbf{HH}}_\ell[i,j] &=& \left\{\begin{array}{ll}0, & |{\mathbf{HH}}_\ell[i,j]|<\tau, \\
             {\mathbf{HH}}_\ell[i,j], & \mbox{otherwise,} \end{array} \right. \nonumber 
\end{eqnarray}
where $0\leq i \leq (M/2)-1$, and $0\leq j\leq (N/2)-1$ (without lost of generality, we assume the image size of $\mathbf{P}$ is $M\times N$ with even $M$ and $N$, otherwise an appropriate image border extension technique can be applied). The hard threshold wavelet denoising technique has been demonstrated to work well even with the very simple Haar wavelet. To extend this scheme to multiple images (${\mathbf P}_1, \ldots, {\mathbf P}_k$), we propose to fuse the low frequency subband images ${\mathbf{LL}}_\ell$ as:
\begin{eqnarray}
	{\mathbf{LL}}=\frac{1}{k} \sum_{\ell=1}^k{\mathbf{LL}}_\ell. 
	\label{LL}
\end{eqnarray}
The high frequency subband images are fused together using the selection of maximum magnitude wavelet coefficients similar to that in \cite{mitra94} together with the hard threshold method such that the wavelet coefficients will be replaced by 0 when the maximum magnitude wavelet coefficients of all the images are smaller than the threshold $\tau$:
\begin{eqnarray}
\mathbf{LH}[i,j] &=& \begin{cases} 0, & \mbox{if} ~\max_\ell |\mathbf{LH}_\ell[i,j]| < \tau, \\
\mathbf{LH}_\ell[i,j], &  \mbox{otherwise;} \end{cases}\notag\\
\mathbf{HL}[i,j] &=& \begin{cases} 0, & \mbox{if} ~\max_\ell |\mathbf{HL}_\ell[i,j]| < \tau, \\
\mathbf{HL}_\ell[i,j], &  \mbox{otherwise;} \end{cases}\notag\\
\mathbf{HH}[i,j] &=& \begin{cases} 0, & \mbox{if} ~\max_\ell |\mathbf{HH}_\ell[i,j]| < \tau, \\
\mathbf{HH}_\ell[i,j], & \mbox{otherwise.} \end{cases} 
\label{LH-HL-HH}
\end{eqnarray}
Together with $\mathbf{LL}$ obtained from \eqref{LL}, the denoised image $\mathbf {\hat X}$ is obtained with the inverse DWT (IDWT) on the subband images $[\begin{array}{llll}\mathbf{LL} & \mathbf{LH} & \mathbf{HL} & \mathbf{HH}\end{array}]$ as:
\begin{eqnarray}
	[\begin{array}{llll}{\mathbf {LL}} & {\mathbf{LH}} & {\mathbf{HL}} & {\mathbf{HH}}\end{array}] \stackrel{IDWT}{\rightarrow} {\mathbf {\hat X}}.
	\label{IDWT}
\end{eqnarray}
With an efficient AWGN denoising method, there leaves us the only hurdle is that the generation of the SAPN denosied image set $({\mathbf{P}}_1,\ldots,{\mathbf{P}}_k)$ from the noise corrupted image $\mathbf{P}$. We propose to apply a two-dimensional version of the RIA method to complete the job.

\subsection{2-D Random Interpolation Averaging} \label{sec:ria}

The RIA method has been demonstrated in \cite{RIA1}  to handle the denosing of AWGN corrupted signal very well. It is, however, the extension of the method  in \cite{RIA1} to 2-D RIA is not obvious. In the paper, we propose to generate a sequence of sub-image $\mathbf P_\ell$ by subsampling with index matrix $\Xi$, such that:
\begin{eqnarray}
	\mathbf \Xi_\ell \in {\mathbb{R}}^{M\times N} \mbox{with } \mathbf \Xi_\ell[i,j]=\{0,1\}
\end{eqnarray}
and hence the sampled sub-image $\mathbf P_\ell$ is given by
\begin{eqnarray}
	\mathbf P_\ell[i,j] = \left\{\begin{array}{ll} \mathbf P[i,j], & \mbox{if } \mathbf \Xi_\ell[i,j]=1, \\
	0, & \mbox{otherwise.}
								\end{array}\right.
\end{eqnarray}
The elements of the index matrix $\mathbf \Xi_\ell[i,j]$ forms a random field.

To ease our discussion, we choose $k=1$, such that there are only 2 sub-images (also because $k=2$ is suffice to show the effectiveness of the algorithm without going into sophisticated generation method for $\mathbf \Xi_\ell[i,j]$). The generation of $\mathbf \Xi_\ell[i,j]$ can be roughly classified as i) non-overlap and ii) overlap. Given two distinct and non-null sampling matrices $\mathbf \Xi_m$ and $\mathbf \Xi_n$  such that $\bigcup_{\ell=1}^k \mathbf \Xi_\ell$ is a matrix will all elements equal 1 that spans the matrix space. These two sampling matrices are non-overlap if and only if $\mathbf \Xi_m[i,j]\cdot \mathbf \Xi_n[i,j]=0$ for all $m\neq n$. These two matrices are said to have overlap if and only if $\mathbf \Xi_m\oplus \mathbf \Xi_n\neq 0$, $\mathbf \Xi_m[i,j]\cdot \mathbf \Xi_n[i,j]\neq 0$ and $(\mathbf \Xi_m[i,j] \neq 0) || (\mathbf \Xi_n[i,j] \neq 0)$ for $m\neq n$. 

In this work, the sub-image sequence will contain two non-overlap sub-images $\mathbf P_0$ and $\mathbf P_1$ and two overlap sub-images $\mathbf P_2$ and $\mathbf P_3$, such that $\mathbf P_0\bigcup \mathbf P_1$ will span $\mathbf P$ and $\mathbf P_2\bigcup \mathbf P_3$ will also span $\mathbf P$. The sampling matrix $\Xi_0$ is generated with exactly $(M\times N)/2$ elements being 0 and others being 1, where the locations of the 0 valued elements within the sampling matrix are uniformly generated in between $[0,M-1]$ and $[0,N-1]$. Furthermore, $\mathbf \Xi_1={\bf{1}}-\mathbf \Xi_0$ with $\bf{1}$ being a matrix with all 1. 

On the other hand, before going on with the generation of $\mathbf \Xi_2$ and $\mathbf \Xi_3$, we need to define three auxiliary non-overlap index matrices $\mathbf \Omega_1, \mathbf \Omega_2 $ and $\mathbf \Omega_3$, such that $\sum_{\ell=1}^3 \mathbf \Omega_\ell[i,j] = 1$ and $\mathbf \Omega_m \cdot \mathbf \Omega_n = 0 $ for all $m \neq n$. Suppose that there is $\eta$ percentage of elements  in $\mathbf \Omega_1$ being 1, and the non-overlap index matrices $\mathbf \Omega_2$ and $\mathbf \Omega_3$ have equal elements of 1, where the locations of 1 inside $\mathbf \Omega_\ell$ are randomly generated with a uniform distribution. The sampling matrices $\mathbf \Xi_2$ and $\mathbf \Xi_3$ are generated by the following expressions:
\begin{eqnarray}
	\mathbf \Xi_2[i,j] = \left\{\begin{array}{ll}  1, & \mbox{if } \mathbf \Omega_1[i,j]=1 ~\mbox{or}~  \mathbf \Omega_2[i,j]=1, \\
	0, & \mbox{otherwise},
								\end{array}\right.
								\label{Xi2}
\end{eqnarray}
\begin{eqnarray}
	\mathbf \Xi_3[i,j] = \left\{\begin{array}{ll}  1, & \mbox{if } \mathbf \Omega_1[i,j]=1 ~\mbox{or}~  \mathbf \Omega_3[i,j]=1, \\
	0, & \mbox{otherwise},
								\end{array}\right.
								\label{Xi3}
\end{eqnarray}
respectively. It is vivid that there is $\eta$ percentage of elements of 1 overlapping in $\mathbf \Xi_2$ and $\mathbf \Xi_3$. Furthermore,  $\mathbf \Xi_2 \bigcup \mathbf \Xi_3 =\mathbf 1$ and $\mathbf \Xi_2 + \mathbf \Xi_3 \neq\mathbf 1$. As an example, consider $\eta = 50\%$, such that there are $(M \times N)/2$ elements being 1 in $\mathbf \Omega_1$, while $\mathbf \Omega_2$ and $\mathbf \Omega_3$ have $(M \times N)/4$ elements being 1. As a result, the sampling matrices $\mathbf \Xi_2$ and $\mathbf \Xi_3$ constructed by \eqref{Xi2} and \eqref{Xi3} will both have $75\%$ elements being 1,  while $\Xi_2$ and $\Xi_3$ only have $50\%$ of matrix entries of 1 overlaps.

\section{SIMULATION RESULTS}

In this section, simulation resuts are carried out to demonstrate the denoising performance of the proposed method (denoted as  "Hard thresholding") on a building ($300 \times 300$), a texture ($240 \times 240$) images and a real image\footnote{It is pictured by Machine Nikon D7000 with resolution $4256 \times 2832$ and we downloaded from https://pixabay.com/en/bricks-brickwork-wall-dirty-1846866/} with large size ($960 \times 638$), compared with several existing image denoising algorithms, i.e., truncated SVD (TSVD) \cite{TSVD}, SALSA \cite{SALSA}, BM3D \cite{BM3D}, OWF \cite{OWF}, VBMFL1 \cite{VBMFL1}. Since our method is developed on the framework of low-rank matrix completion, TSVD is intuitively considered as a benchmark, where it is widely used in matrix completion for its advantages of easy implementation and effectively denoising Gaussian noise. What is more, the solution of TSVD is obtained by a filtering of the singular value decomposition: the components corresponding to singular values much larger than a threshold parameter are taken without any significant modification, whereas the components corresponding to singular values much smaller than the threshold parameter are essentially removed. It is similar with the strategy of the hard threshold wavelet denoising method used in our method, and that is why we want to compare with TSVD in our work. We shall also include the control experiment of simple averaging (denoted as "Average") being applied to the 2-D RIA LRMF-AM result $\mathbf{\hat P}_\ell$, such that the denoised image is obtained as $\mathbf {\hat X} = {1\over k} \sum\limits_{\ell=1}^k {\mathbf {\hat P}_\ell}$.

Two objective measures, namely, PSNR and SSIM \cite{SSIM} indexes, are adopted to provide quantitative and quality evaluations of the denoising results. For the ground truth image $\mathbf X$ and denoised image $\mathbf {\hat X}$, the PSNR is defined as:
\begin{align}
{\rm PSNR} = 10 {\rm log}_{10}\frac{255^2}{\rm MSE}
\end{align}
where $255$ is the peak value of the gray-scale image $\mathbf X$ under concern, with the mean squares error (MSE) given by
\begin{align}
{\rm MSE} = \frac{1}{MN}\sum_{i = 1}^M\sum_{j = 1}^N|\mathbf X[i,j] - \mathbf {\hat X}[i,j]|^2.
\end{align}
Therefore, the smaller the MSE, the larger the PSNR, which implies the better denoising. Since the PSNR is often inconsistent with human eye perception, even though it is the mostly used quality measure, the SSIM is employed to comprehensively reflect the performance of the denoising methods. Our method produces results superior to those of most methods in both visual image quality and quantitative measures.  $\mathbf {\hat X}$ is computed based on 200 independent Monte Carlo trials of generating random noise (AWGN and SAPN) corrupted images.

The noise corrupted images considered in our simulation are obtained with strong AWGN with standard deviation $\sigma = 20$ and the normalized noise intensity $\rho$ of SPAN being $\rho = 0.3 (\rho = p_1 + p_2)$. There are four algorithmic parameters of our methods, namely, the percentage of overlapped sub-images pixels $\eta $, the number of non-overlapped sub-images, rank $r$ and hard thresholding parameter $\tau$. It is necessary to discuss how to choose them appropriately.  It can  be selected by empirical results from $a$ $priori$ research results obtained from wavelet hard threshold denoising. Note that for the method of "Average", it does not need to determine the hard thresholding parameter $\tau$.  In this work, it is worth noting that although the threshold value $\tau$ can be obtained by choosing the $(r+1)$th singular value of the noisy image, our simulation results indicate that fixing it at around $\tau = 15$ is good enough for gray-level images. 
\subsection{Effect of algorithmic parameter selection}
\begin{figure}[!htbp]
\begin{center}
\includegraphics[width=8.5cm]{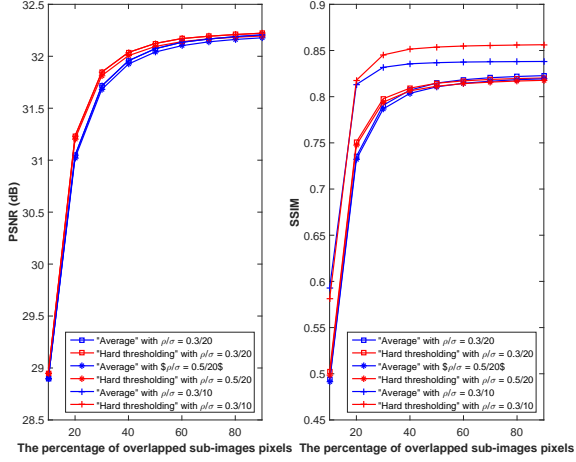}
\end{center}
\caption{Effect of percentage of overlapped sub-images pixels, i.e., $\eta$, where there is $\eta$ percentage of elements of 1 overlapping in $\mathbf \Xi_2$ and $\mathbf \Xi_3$, and two non-overlapped sub-images are builtd based on the definition of $\mathbf \Xi_0$ and $\mathbf \Xi_1$.}
\label{main2-1}
\end{figure}
\begin{figure}[!htbp]
\begin{center}
\includegraphics[width=8.5cm]{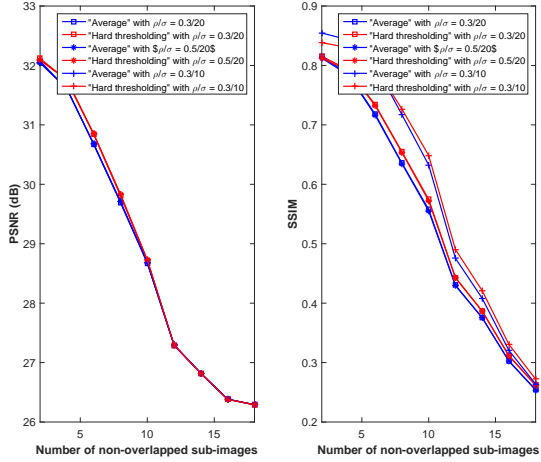}
\end{center}
\caption{Effect of non-overlapped sub-image number, where two overlapped sub-images are constructed based on the definition of $\mathbf \Xi_2$ and $\mathbf \Xi_3$ with $\eta = 50\%$.}
\label{main021}
\end{figure}

Fig. \ref{main2-1} plots the PSNR and SSIM versus the percentage of overlapped sub-images pixels in different ratios of mixed noise, including $\rho/\sigma = 0.3/20$, $\rho/\sigma = 0.5/20$ and $\rho/\sigma = 0.3/10$. To clearly show the effect of the percentage of overlapped sub-images pixels, we consider that the number of non-overlapped sub-images is two. When the percentage of overlapped sub-images pixels is larger than $\eta = 40\%$, both the PSNR and SSIM indexes almost keep stable. Therefore, in the following simulations, the percentage of overlapped sub-images pixels is set as $\eta = 50\%$.

Fig. \ref{main021} plots the PSNR and SSIM versus the number of non-overlapped sub-images in different ratios of mixed noise.  We see that both the PSNR and SSIM of the proposed methods (i.e., "Average" and "Hard thresholding") decrease with the increase of non-overlapped sub-images number, and our methods attain good noise removal when the number is 2. It is preferred to select the number of non-overlapped sub-images as 2 for the following simulations. Thus, the method of "Hard thresholding" outperforms that of "Average" for image denoising in small number of non-overlapped sub-images.
\begin{figure}[!htbp]
\begin{center}
\includegraphics[width=8.5cm]{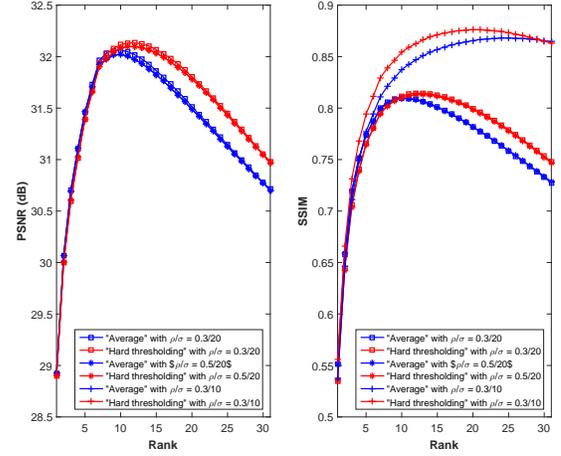}
\end{center}
\caption{Effect of the proposed method with rank $r$ varying from $1$ to $31$, based on two non-overlapped sub-images and two overlapped sub-images with $\eta = 50\%$. The standard deviation $\sigma$ of AWGN and the normalized noise intensity $\rho$ of SPAN are set at $\sigma = 20$ and $\rho = 0.3$, respectively.}
\label{main3}
\end{figure}
\begin{figure}[!htbp]
\begin{center}
\includegraphics[width=8.5cm]{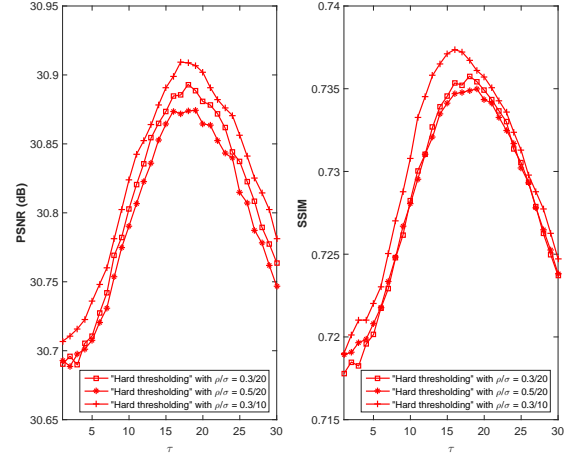}
\end{center}
\caption{Effect of hard-thresholding parameter $\tau$ varying from 2 to 20, based on two non-overlapped sub-images and two overlapped sub-images with $\eta = 50\%$. The rank $r = 10$, $\sigma =20$ and $\rho = 0.3$.}
\label{main4}
\end{figure}

In Fig. \ref{main3}, the effect of rank $r$ on the denoising performance in different ratios of mixed noise is investigated in the sampled grids from $1$ to $31$ with interval 2. The PSNR and SSIM are considered as the metric to determine the optimal rank, respectively. It is seen that the highest PSNR and SSIM are obtained when rank $r = 10$, which implies that our methods have achieved the best image denoising performance at $r = 10$. Again, our method with simple averaging fusion is inferior to one with hard thresholding fusion in terms of both PSNR and SSIM.
\begin{figure}[!htbp]
\begin{center}
\includegraphics[width=8.5cm]{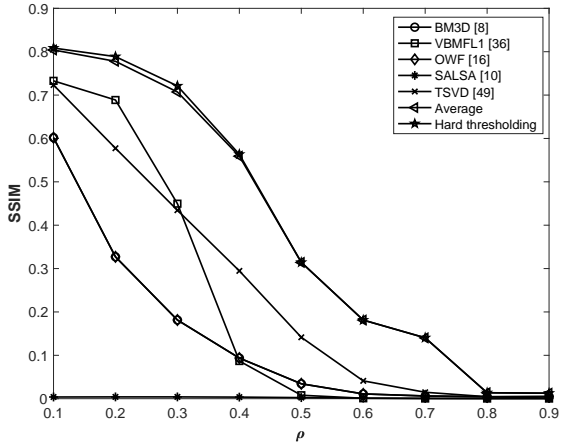}
\end{center}
\caption{SSIM versus impulsive noise level ranging from $0.1$ to $0.9$, where the standard deviation of AWGN is fixed at $\sigma = 20$. The rank $r = 10$ and $\tau = 15$.}
\label{main7}
\end{figure}

\begin{figure*}[!htbp]
\begin{center}
\includegraphics[width=17cm]{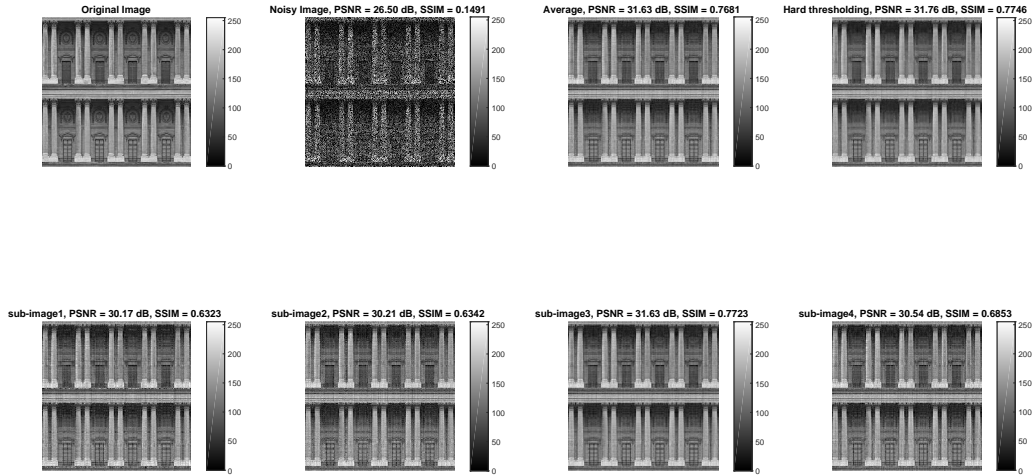}
\end{center}
\caption{Performance comparison of each sub-image}
\label{main11}
\end{figure*}

Fig. \ref{main4} plots the PSNR and SSIM versus different hard thresholding parameter $\tau$ of denoising a building image in different ratios of mixed noise. Although a popular method to determine the optimal value is using the statistical information of noise, it it impractical because the standard deviation $\sigma$ needs to be known. Moreover, when the image is corrupted by more complicated noise distribution, e.g., mixed noise, it is difficult to choose the optimal value based on the statistical information of noise. Therefore, we select the hard thresholding parameter $\tau$ via $(r+1)$th largest singular value $\lambda_{r+1}$. Assume that both refer to singular values are arranged in the descending order. At first, we test the denoising performance with different grids of the $(r+1)$ singular value, where  $\lambda_{r+1}$ is sampled with interval grid 1 from 0 to the value of $(r+1)$th singular value. $\tau$ has a wide range from $\tau = 11$ to $\tau = 17$ that generates faithful denoising results, where the differences of the PSNR and SSIM in the denoised image are less than $0.030$ dB and $0.001$, respectively. The optimal results are achieved at $\tau = 15$, which makes the selection of $\tau$ robust to user. We observe that when $\tau \approx 15$ $, \lambda_{r+1} \approx 30$, i.e., $\tau \approx \lambda_{r+1}/2$. Therefore, the hard thresholding parameter $\tau$ of our method "Hard thresholding" is adopted at $\tau = \lambda_{r+1}/2$ for the following simulation results.  The threshold $\tau_{universal} = \sigma\sqrt{2log(L)}$ ($L = M \times N$ is the number of pixels in image) is well known in wavelet literature \cite{HOSVD, RWBSRR, AUSWS} as the $Universal$ $threshold$, which is the optimal threshold in the asymptotic sense. Compared with the universal threshold, the best empirical threshold $\tau = 15$ adopted in our work is much lower than  $\sigma\sqrt{2log(L)}$, which indicates that the proposed method is not sensitive to the threshold value, as a wide range of threshold can be selected especially for large matrix size. Since the "Average" method is insensitive to the hard-thresholding parameter $\tau$, it is not included in Fig. \ref{main4}, where the PSNR and SSIM are 31.99 dB and 0.7456, respectively.

\subsection{Performance Comparison}
Unless stated otherwise, two non-overlapped sub-images in the first set, two partially overlapped sub-images in the second set with $\eta = 50\%$ pixels, the rank $r = 10$ and the hard thresholding parameter $\tau = 15$ for "Hard thresholding" method and $\sigma =20$, $\rho = 0.3$ are adopted.

\begin{figure*}[!htbp]
\begin{center}
\includegraphics[width=17cm]{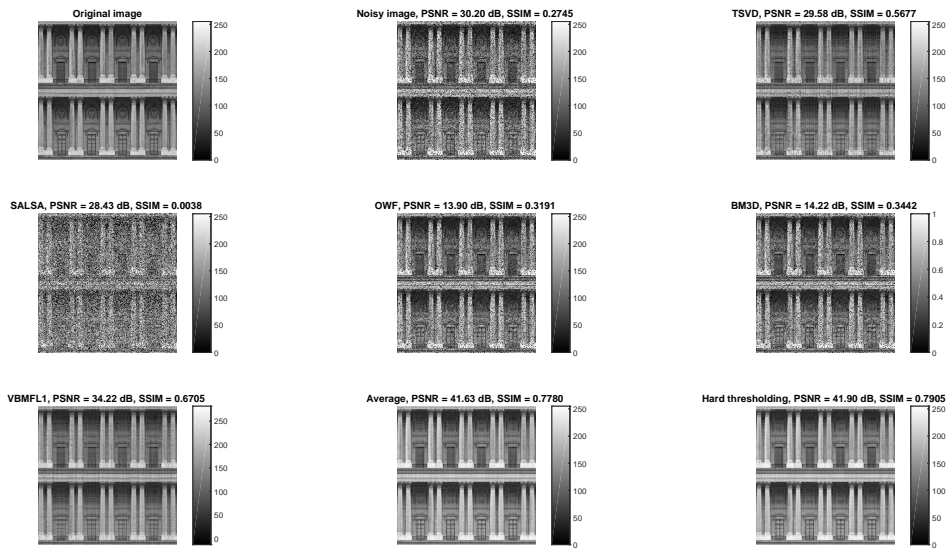}
\end{center}
\caption{Performance comparison of a building image in mixed noise.}
\label{main01}
\end{figure*}

Fig. \ref{main7} plots the curves of SSIM versus the impulsive noise level among different algorithms in the mixed noise. For the comparison of SSIM, we observe that "Hard thresholding" method yields the best performance on denoising a building image with the SAPN to AWGN ratio ranging from $0.1$ to $0.9$, where the standard deviation of AWGN is fixed at $\sigma = 20$.

Fig. \ref{main11} compares the performance of each restored sub-image before fusion operator, where sub-images 1 and 2 are restored from the first image set, sub-images 3 and 4 are recovered from the second image set. After the fusion operation, we have found that both PSNR and SSIM of the combined image are higher than that of each sub-image.

Fig. \ref{main01} shows the problem of denoising a building image from the mixed noise. It is observed that our methods, TSVD and VBMFL1 give satisfactory results while BM3D, SALSA and OWF fail in denoising in the presence of the mixed noise. SALSA becomes worse in the mixed noise in terms of the PSNR and SSIM. It makes sense because it is not robust against the mixed noise especially with a strongly Gaussian noise although it can achieve faster computation. Interestingly, to some extent, TSVD is effective to eliminate the noise since the denoising principle of TSVD is to abandon the information of remaining smaller singular values (regarded as the noise information). Thus, we have found that VBMFL1 also yields satisfactory result similar to our methods with the help of the constructed hierarchical Bayesian generative model. However, without using $a$ $priori$ information of the mixed noise, the denoising performance may degrade substantially. In this sense, our methods enjoy more applicability compared with these existing algorithms. Moreover, our methods achieve the best denoising performance among all compared algorithms, and the method "Hard thresholding" is slightly superior to the "Average" method.

\begin{figure*}[!htbp]
\begin{center}
\includegraphics[width=17cm]{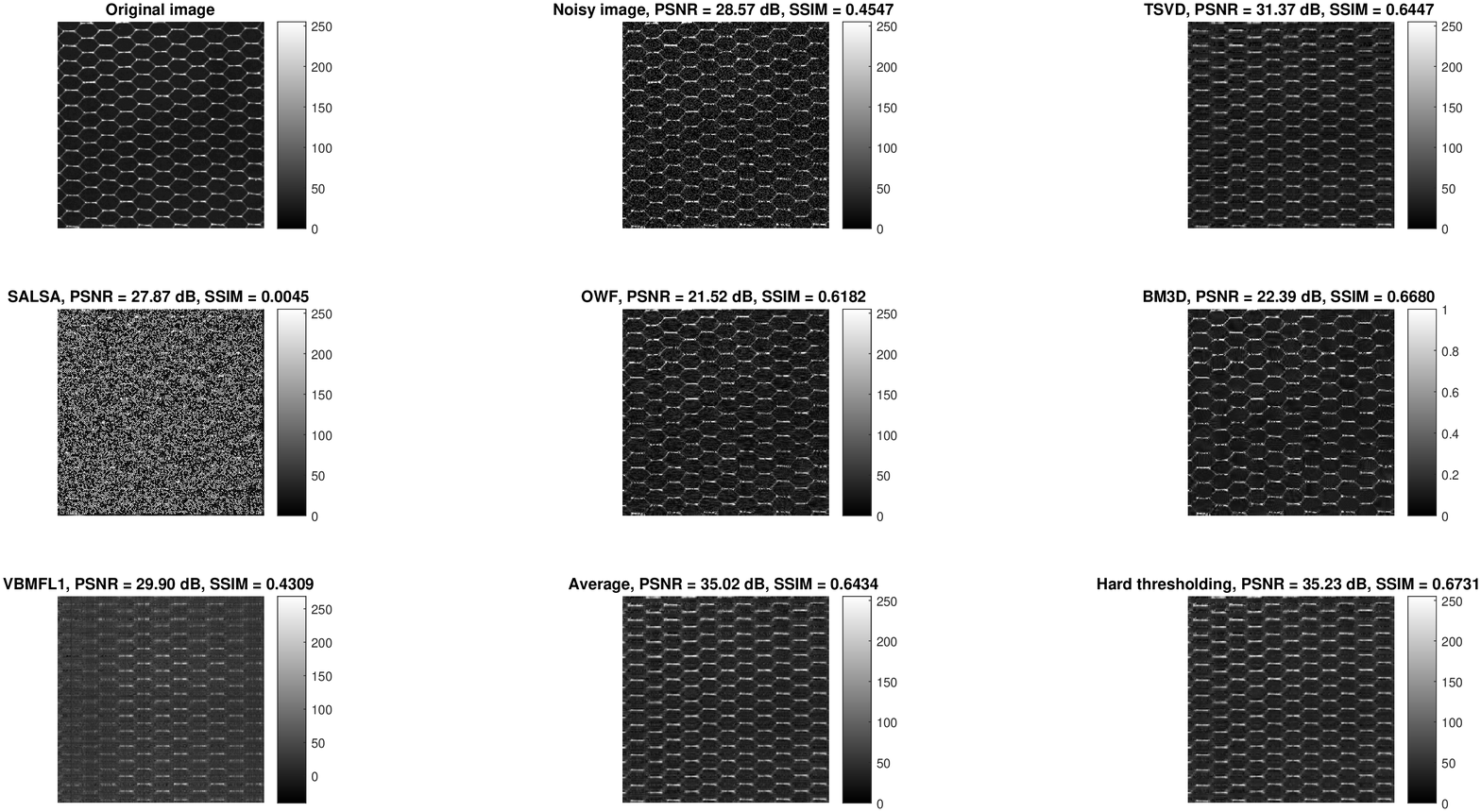}
\end{center}
\caption{Performance comparison of a texture image in mixed noise.}
\label{main8}
\end{figure*}

Fig. \ref{main8} tests the denoising performance of a texture image among different algorithms. It is observed that VBMFL1 fails to restore the texture image from mixed noise, although it has good denoising on building image. Still, SALSA cannot work on the situation of mixed noise. Thus, consistent with the results of a building image, our methods achieve the highest PSNR on a texture image, while they enjoy comparable SSIM compared with TSVD and BM3D, which implies that they can also preserve the image structure well.

\begin{figure}[!htbp]
\begin{center}
\includegraphics[width=9cm]{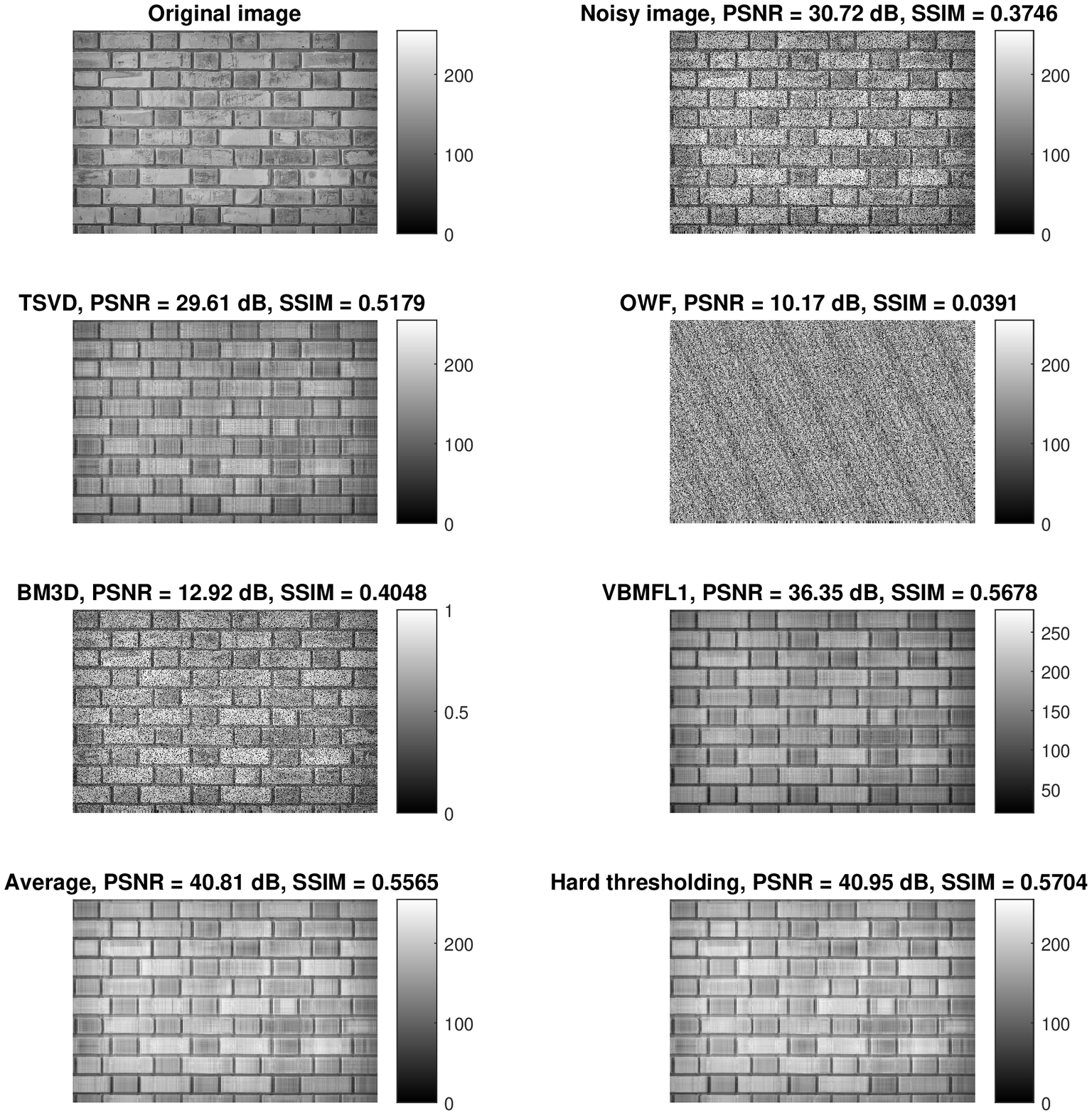}
\end{center}
\caption{Performance comparison of a large-size image in mixed noise}
\label{main10}
\end{figure}

To further investigate the denoising performance of the proposed methods, simulation result on a real image with large size is implemented, shown in Fig. \ref{main10}.  TSVD, BM3D and OWF are not robust to the mixed noise for restoring the ground truth image with large size, while VBMFL1 and our methods have better denoising performance. Compared with the VBMFL1, the proposed methods enjoy comparable performance in terms of SSIM, and higher PSNR.

\section{CONCLUSIONS}

In this paper, a new and effective denoising approach has been developed to preserve the spatial image structure based on the basic framework of LRMF-AM. Although the denoising results generated by LRMF-AM  for the mixed noise (SAPN and AWGN) are generally satisfactory, the reconstructed images are still contaminated by AWGN in the smooth areas of the images. Herein, 2-D RIA has been taken advantage under the benefits of randomly generated sub-image sequence, including non-overlap and overlap sub-images. What is more, the mean filtering effect of the image fusion process via a hard threshold wavelet denoising method is utilized to suppress the AWGN noise as the result of its independent nature. The superior performance of the proposed approach is validated by extensive simulation results. It is expected that the proposed method will be applicable to general noisy image quality enhancement and also to computer vision problems.

\bibliographystyle{IEEEtran}
\bibliography{IEEEabrv}

\end{document}